\documentclass[fleqn,twoside]{article}
\usepackage{espcrc2}

\newcommand{\AmS}{{\protect\the\textfont2
  A\kern-.1667em\lower.5ex\hbox{M}\kern-.125emS}}

\title{\textbf{Remarks on monopoles in Abelian projected continuum Yang-Mills
theories\footnote{Contribution to LHP2001, 9-18 July 2001, Cairns, Australia
}}}

\author{A. R. Fazio\address[MCSD]{Universit{\`{a}} degli Studi di Milano and INFN, \\ 
via Celoria 16, 20143 Milano, Italy}
S.P. Sorella\address{UERJ, Universidade do Estado do Rio de
Janeiro\\
Rua S\~{a}o Francisco Xavier 524, 20550-013 Maracan\~{a},
Rio de Janeiro, Brazil}}
\begin{document}

\begin{abstract}
A possible mechanism accounting for monopole configurations in continuum
Yang-Mills theories is discussed. The presence of the gauge fixing term is
taken into account.
\end{abstract}

\maketitle

\section{Introduction}

The understanding of confinement in nonabelian gauge theories is one of the
major challenge in theoretical physics. The idea that confinement could be
explained as a dual Meissner effect for type II superconductors is largely
accepted, with confirmations from lattice simulations.

A key ingredient for the picture of dual superconductivity is the mechanism
of Abelian projection introduced by 't Hooft \cite{thooft}, which consists
of reducing the gauge group $SU(N)$ to an Abelian subgroup, identified with
the Cartan subgroup $U(1)^{N-1}$, by means of a partial gauge fixing. This
is achieved by choosing any local composite operator $X(x)$ which transforms
in the adjoint representation, $X^{^{\prime }}(x)=UX(x)U^{\dag }.$ The gauge
is partially fixed by requiring that $X$ becomes diagonal, $X^{^{\prime
}}(x)=\mathrm{diag}(\lambda _1(x),......,\lambda _N(x)),$ where $\lambda
_i(x)\;$denote the gauge invariant eigenvalues. As shown in \cite{thooft},
monopoles configurations appear at the points $x_0$ of the space-time where
two eigenvalues coincide, \textit{i.e.} $\lambda _{i+1}(x_0)=\lambda
_i(x_0).\;$Further, the gauge field is decomposed into its diagonal and
off-diagonal parts. The diagonal components correspond to the generators of
the Cartan subgroup and behave as photons. The off-diagonal components are
charged with respect to the Abelian residual subgroup and may become massive 
\cite{m1,m2}, being not protected by gauge invariance. This mass should set
the confinement scale, allowing for the decoupling of the off-diagonal
fields at low energy. The final Abelian projected theory turns out thus to
be described by an effective low-energy theory in which the relevant degrees
of freedom are identified with the diagonal components of the gauge fields
and with a certain amount of monopoles, whose condensation should account
for the confinement of all chromoelectric charges. Lattice simulations \cite
{kronf,suzuki} have provided evidences for the Abelian dominance hypothesis,
according to which QCD\ in the low-energy regime is described by an
effective Abelian theory. This supports the realization of confinement
through a dual Meissner effect, although the infrared Abelian dominance in
lattice calculations seems not to be a general feature of any Abelian gauge 
\cite{dig}. Furthermore, many conceptual points remain to be clarified in
order to achieve a satisfactory understanding of confinement in the
continuum. Certainly, the problem of the derivation of the Abelian dominance
from the QCD Lagrangian is a crucial one. Also, the characterization of the
effective low-energy Abelian projected theory and of its monopoles content is
of great relevance. There, one usually starts by imposing the so called
Maximal Abelian Gauge (MAG) \cite{ch}, which allows for a manifest residual
subgroup $U(1)^{N-1}$. The presence of monopoles in the MAG follows then
from $\Pi _2(SU(N)/U(1)^{N-1})=Z^{N-1}$. However, being the MAG\ a
gauge-fixing condition, it is manifestly noncovariant. Therefore, monopoles
here do not seem to be directly related to the singularities occurring for
coinciding eigenvalues in the process of diagonalization of a local
covariant operator $X(x).$ Rather, they are associated to singular
configurations of the fields \cite{ch}.

The purpose of this contribution is to discuss a possible mechanism
accounting for the presence of monopoles in the MAG, for continuum gauge
theories. The argument turns out to be generalized to any renormalizable
gauge, the main idea being that of showing that 't Hooft Abelian projection
can be suitably carried out in the presence of gauge fixing terms.

\section{Monopoles in quantized Yang-Mills theories}

In what follows we present a simple way in order to account for monopoles in
continuum quantized Yang-Mills theories. In particular, we point out that it
is possible to introduce in the path integral a covariant local quantity
whose diagonalization is compatible with the gauge fixing, reproducing at
the end the usual form of the Yang-Mills partition function in the presence
of monopoles \cite{ch}.

Let us start by considering the partition function for the quantized $SU(N)$
Yang-Mills theory
\begin{eqnarray}
Z&=&{\mathcal N}\int [D\Phi ][DA]\exp \Bigg( -\int d^4x\mathrm{Tr}\frac
14F_{\mu \nu }F_{\mu \nu }\nonumber\\
&&-S_{\mathrm{GF}}(A,b,c,{\bar{c})}\Bigg) 
\end{eqnarray}
where $S_{\mathrm{GF}}$ denotes the gauge-fixing action including the
Faddeev-Popov ghosts. We do not specify further the term $S_{\mathrm{GF}}$,
which can be any renormalizable gauge fixing action as, for instance, the
MAG condition, the Landau gauge, etc. The measure $[D\Phi ]$ denotes
integration over the Lagrange multiplier $b$ and the ghost fields $c,\bar{c}$%
.

We proceed by rewriting the term $\mathrm{Tr}F_{\mu \nu }F_{\mu \nu }$ in a
first order formalism by introducing an antisymmetric two-form field $B_{\mu
\nu }$ \cite{fm,ko}
\begin{equation}
\mathrm{Tr}\frac 14F_{\mu \nu }F_{\mu \nu }\rightarrow \mathrm{Tr}\left(\frac
i2F_{\mu \nu }B_{\mu \nu }\right)+\mathrm{Tr}\left(\frac 14B_{\mu \nu }B_{\mu \nu }\right)\
\; 
\end{equation}
Therefore, for the partition function we get
\begin{eqnarray}
Z &= &{\mathcal N}\int [D\Phi ][DA][DB]  \exp\Bigg[ - S_{\mathrm{GF}} \nonumber\\
&& -\int d^4x\mathrm{Tr}\left(
 \frac {iF_{\mu \nu }B_{\mu \nu }}2+\frac {B_{\mu \nu }B_{\mu \nu }}4\right) \Bigg] 
\end{eqnarray}
Notice that the field $B_{\mu \nu }$ transforms covariantly
under a gauge transformation of $SU(N)$
\[
B_{\mu \nu }\longrightarrow B_{\mu \nu }^U=UB_{\mu \nu }U^{\dag }\ , 
\]
from which it follows that the quadratic term $\mathrm{Tr}B_{\mu \nu }B_{\mu
\nu }$ is left invariant
\[
\mathrm{Tr}B_{\mu \nu }B_{\mu \nu }=\mathrm{Tr}B_{\mu \nu }^UB_{\mu \nu
}^U\;. 
\]
Also, it is worth remarking that the field $B_{\mu \nu }$ does not appear in
the gauge fixing term $S_{\mathrm{GF}}(A,b,c,{\bar{c})}$. According to 't
Hooft procedure, we can now pick up any component of $B_{\mu \nu },$ say $%
B_{12}$, and, due to its hermiticity, diagonalize it by a suitable transformation $\Omega $ of $SU(N)$, namely 
\[
B_{12}\rightarrow B_{12}^{diag}=\Omega B_{12}\Omega ^{\dag }\;. 
\]
Due to the invariance of $\mathrm{Tr}B_{\mu \nu }B_{\mu \nu },$ we have
\begin{eqnarray}
\mathrm{Tr}B_{\mu \nu }B_{\mu \nu } &=&\mathrm{Tr}\left(
2B_{12}B^{12}+B_{jk}B^{jk}\right)  \nonumber  \label{ind} \\
&=&\mathrm{Tr}\left( 2\Omega B_{12}\Omega ^{\dag }\Omega B^{12}\Omega
^{\dag }\right.\nonumber  \label{ind}\\&&\left.+\Omega B_{jk}\Omega ^{\dag }\Omega B^{jk}\Omega ^{\dag }\right) 
\nonumber \\
&=&\mathrm{Tr}\left( 2B_{12}^{diag}B_{12}^{diag}\right.\nonumber \label{ind}\\&&\left.+\Omega B_{jk}\Omega
^{\dag }\Omega B^{jk}\Omega ^{\dag }\right) \;,  \label{ind}
\end{eqnarray}
where the sum over the indices $(j,k)$ does not include the component $%
B_{12}.$ The partition function $Z$ becomes 
\begin{eqnarray*}
Z &=&{\mathcal N}\int [D\Phi ][DA][DB][D\Omega ]  \exp \Bigg\{ -S_{%
\mathrm{GF}} \\
&& -\int d^4x\Bigg[\frac i2\mathrm{Tr}F_{\mu \nu }B_{\mu \nu }  \\
&&+\frac 14%
\mathrm{Tr}\left( 2\Omega B_{12}\Omega ^{\dag }\Omega B^{12}\Omega ^{\dag
}+\Omega B_{jk}\Omega ^{\dag }\Omega B^{jk}\Omega ^{\dag }\right)\Bigg]\Bigg\}
\end{eqnarray*}
where we have inserted the integration measure $[D\Omega ]$ over the gauge
transformations which diagonalize $B_{12}$. This is always possible, thanks
to eq.$\left( \ref{ind}\right) $. Performing now the change of variables
\begin{equation}
B_{\mu \nu }\rightarrow \Omega ^{\dag }B_{\mu \nu }\Omega \ ,\quad \Omega
\rightarrow \Omega \ ,  \label{ch}
\end{equation}
we obtain
\begin{eqnarray*}
Z&=&{\mathcal N}\int [D\Phi ][DA][DB][D\Omega ]\nonumber\\
&&\exp \Bigg(\int d^4x\mathrm{Tr}%
\Bigg[-\frac i2\Omega F_{\mu \nu }\Omega ^{\dag }B_{\mu \nu }\nonumber\\
&&-\frac 14B_{\mu \nu }B_{\mu \nu }\Bigg] -S_{\mathrm{GF}}\Bigg)
\end{eqnarray*}
The change of variables $\left( \ref{ch}\right) $ has the effect of moving
the $\Omega $'s from the quadratic term $BB$ to the first term $FB$.
Recalling then that the $\Omega $'s are precisely those transformations
which diagonalize $B_{12}$, it follows that
\begin{eqnarray*}
\Omega F_{\mu \nu }\Omega ^{\dag } &=&\Omega \Bigg(\partial _\mu A_\nu
-\partial _\nu A_\mu -\left[ A_\mu ,A_\nu \right]\\ 
&&+([\partial _\mu ,\partial
_\nu ]\Omega ^{\dag })\Omega\Bigg) \Omega ^{\dag },\\
F_{\mu \nu } &=&\left[ D_\mu ,D_\nu \right] \ .
\end{eqnarray*}
Finally, we can path integrate the field $B$ obtaining the expression
\begin{eqnarray*}
Z&=&{\mathcal N}\int [D\Phi ][DA][D\Omega]\\
&&\exp \int d^4x\Bigg(Tr\left[ -\frac
14(F_{\mu \nu }^{reg}+F_{\mu \nu }^{sing})^2\right]
-S_{\mathrm{GF}}\Bigg)  
\end{eqnarray*}
with
\[
F_{\mu \nu }^{sing}=([\partial _\mu ,\partial _\nu ]\Omega ^{\dag })\Omega
\;. 
\]
Notice that $F_{\mu \nu }^{sing}$ is nonvanishing when the transformation $%
\Omega $ is singular. This occurs for coinciding eigenvalues of the $B_{12}$%
. Needless to say, these singularities correspond to monopoles. The sum over
the singular transformations may be represented as an integration over
the surfaces corresponding to the closed loop currents of the monopoles \cite
{ch}.

We end up thus with the usual form of the Yang-Mills partition function in
which monopole configurations appear explicitly. Of course, the introduction
of $B_{\mu \nu }$ has to be regarded as a trick for inserting in the path
integral a covariant field which can be diagonalized, according to 't Hooft
procedure. It is apparent that the field $B_{\mu \nu }$
represents indeed the field strength $F_{\mu \nu }.$ The meaning of
this procedure is that we should be able to introduce monopole
configurations in the expression for the partition function, regardless of
the particular gauge fixing adopted.

\section{Acknowledgements}

We are grateful to V. Bornyakov, M. Polykarpov and  V. I. Zakharov,  for enlightening discussions on the subject. S.P.~Sorella thanks the Theoretical Physics
Department of the Milano University (Italy) and the INFN (sezione di Milano)
for the kind invitation.The Conselho Nacional de Desenvolvimento
Cient\'{\i}fico e Tecnol\'{o}gico CNPq-Brazil, the Funda{\c {c}}{\~{a}}o de
Amparo {\`{a}} Pesquisa do Estado do Rio de Janeiro (Faperj), the SR2-UERJ,
the INFN (sezione di Milano) and the MIUR, Ministero dell'Istruzione
dell'Universit\`{a} e della Ricerca, Italy, are acknowledged for the
financial support.

\end{document}